\documentclass[conference,compsoc]{IEEEtran}
\usepackage{cite}
\usepackage{amsmath,amssymb,amsfonts}
\usepackage{graphicx}
\usepackage{textcomp}
\usepackage{amsmath}
\usepackage{graphicx}
\usepackage{hyperref}
\usepackage{url}
\usepackage{array}
\usepackage{listings}
\usepackage[noend]{algpseudocode}
\usepackage{caption}
\usepackage{subcaption}

\usepackage{booktabs}
\usepackage{tabularx} 
\makeatletter
\g@addto@macro{\UrlBreaks}{\UrlOrds}
\makeatother
\def\BibTeX{{\rm B\kern-.05em{\sc i\kern-.025em b}\kern-.08em
    T\kern-.1667em\lower.7ex\hbox{E}\kern-.125emX}}
\begin{document}

\title{Vulnerabilities of smart contracts and mitigation
schemes: A Comprehensive Survey}


\author{\IEEEauthorblockN{Wejdene Haouari}
\IEEEauthorblockA{Department of Electrical Engineering\\ \& Computer Science, \\York University, ON, Canada}
\and
\IEEEauthorblockN{Abdelhakim Senhaji Hafid}
\IEEEauthorblockA{Department of Computer Science \\and Operational Research, \\
 University of Montreal, QC, Canada}
\and
\IEEEauthorblockN{Marios Fokaefs}
\IEEEauthorblockA{Department of Electrical Engineering\\ \& Computer Science, \\York University, ON, Canada}}

\maketitle

\begin{abstract}
Ethereum smart contracts are highly powerful; they are immutable and retain massive amounts of tokens. However, smart contracts keep attracting attackers to benefit from smart contract
flaws and Ethereum’s unexpected behavior. Thus, methodologies and tools have been proposed to help implement secure smart contracts and to evaluate the security of smart contracts already deployed.
Most related surveys focus on tools without discussing the logic behind them; in addition, they assess the
tools based on papers rather than testing the tools and collecting community feedback. Other surveys lack
guidelines on how to use tools specific to smart contract functionalities.
This paper presents a literature review combined with an experimental report that aims to assist developers
in developing secure smarts, with a novel emphasis on the challenges and vulnerabilities introduced by
NFT fractionalization by addressing the unique risks of dividing NFT ownership into tradeable units called fractions. It provides a list of frequent vulnerabilities and corresponding mitigation solutions. In addition, it evaluates the community’s most widely used tools by executing and testing them on sample smart contracts.
Finally, a comprehensive guide on how to implement secure smart contracts is presented.
\end{abstract}


\section{Introduction}
Blockchain technology has exploded in popularity over the years due to its immutability, security, and transparency in a permissionless and decentralized environment. One of the well-known and used implementations is Ethereum. As of 31 May 2023, the Ethereum cryptocurrency's market capital surpassed \$220 billion, with millions of transactions being executed every day \cite{etherscan}. 
Ethereum uses Turing Complete blockchain technology \cite{ethereumwhitepaper}, allowing developers to implement smart contracts. Smart contracts are executable programs stored in the blockchain written primarily in solidity \cite{Solidity}. Using smart contracts facilitates the execution of pre-defined terms without consulting third parties in an anonymous, transparent, and tempered manner.

The Ethereum blockchain is the foundation for numerous financial projects. Examples include (a) platforms for decentralized finance (DeFi), which enable users to access various financial services like lending and borrowing, trading, and insurance without intermediates. (b) Stablecoins, cryptocurrencies whose value is tied to that of a fiat currency or other asset, like gold or the US dollar; (c) Security token offerings (STOs), which are virtual assets that simulate ownership of a physical asset like a stock or piece of real estate. (d) Non-fungible tokens (NFTs), which are unique digital assets that might signify ownership of a physical or digital object, like a work of art or a collectible, (e)  Amidst this variety, NFT fractionalization emerges as a noteworthy trend, dividing ownership of NFTs into smaller, more accessible units, thus broadening participation in the digital asset market. This value attracts attackers to exploit different vulnerabilities related to implementing smart contracts to steal cryptocurrencies or tamper with assets. For instance, a recent attack was on May 2023 where Level Finance Exchange announced the loss of more than 214,000 \$LVL tokens, approximated to be \$1.01 million \cite{LevelFinance}, caused by an attack on its smart contract by manipulating a recursive calling vulnerability. 
There are various vulnerabilities and causes, such as arithmetic overflows, reentrancy, inadequate randomness, calling an unknown third-party contract code, and failing to check the return status of external calls. The runtime environment that supports contract code execution provides more attack points. One example is when malevolent miners pick and choose which transactions are included in a mined block or in what order they are included.

Since vulnerabilities in smart contracts can result in significant financial loss, multiple tools are proposed to identify vulnerabilities in Solidity smart contracts. Detection methods include (a) Symbolic execution when the program is abstractly executed to cover various possible inputs; (b) Fuzzy testing by injecting several invalid inputs; (c)Taint analysis by checking an input flow; and (d) Formal verification, when the behavior of the smart contract is checked mathematically using a formal model.

In this paper, we conduct a literature review and an experimental report.  
We present the most common vulnerabilities in Solidity smart contracts, including those posed by the fractionalization of NFTs and the corresponding mitigation schemes. We also overview and compare the most popular schemes and tools used to detect smart contract vulnerabilities. Finally, we propose a set of guidelines for auditing smart contracts.

In this survey, we searched 2 popular databases, Engineering Village and Scopus, and web articles written by the Ethereum community Concerning vulnerabilities in Solidity smart contracts, mitigation tools, and guidelines. Moreover, we experimentally studied 5 tools widely used to detect vulnerabilities in smart contracts, namely Oyente \cite{oyente}, Slither \cite{slither}, Mythril  \cite{mythril}, Manticore \cite{manticore} and Echidna \cite{echidna}.  

The rest of this paper is organized as follows: Section \ref{methodology} outlines the methodology we used to produce this survey. Section \ref{vul} presents the most common vulnerabilities of smart contracts and the corresponding mitigation schemes. Section \ref{Detection} presents standard methodologies for detecting smart contract vulnerabilities. Section \ref{Vulnerabilities} introduces the most valuable tools to detect vulnerabilities in smart contracts. Section \ref{dis} presents a set of guidelines for auditing smart contracts. Section \ref{related}  presents an overview of existing related surveys. Finally, Section \ref{conc} concludes the paper.

\section{Survey Methodology }
\label{methodology}
In this section, we present the primary research questions that we hope to answer through this survey. The methodology we used to collect existing related work is then presented.

This survey aims to answer the following research questions: 
\begin{itemize}
    \item RQ1: What are the most frequent vulnerabilities in Solidity smart contracts? 
    \item RQ2: How to mitigate vulnerabilities in Solidity smart contracts? 
    \item RQ3: What are the existing methodologies used to detect the vulnerabilities of smart contracts, and how do they compare? 
    \item RQ4: what are the most used tools by the Ethereum community to investigate/mitigate smart contracts based on Github forks and web articles?
\end{itemize}

To address these questions, we conducted a Multivocal Literature Review (MLR)\cite{10.1145/2915970.2916008} for data preparation; 
 This technique incorporates both gray literature and white literature. Gray literature includes blogs, videos, and forums; it is usually written by practitioners from industry and academia. It's not peer-reviewed. White literature contains peer-reviewed research articles from journals. We
chose the published studies in
journals and conferences with high-impact factors and competitive
acceptance rates. We also check the citation count of the studies
being chosen on Google Scholar to evaluate their impact on
the evolution of this emerging paradigm.
The rationale behind choosing MLR is that the field of smart contracts security is still relatively new; thus, including literature from practitioners helps provide a complete overview. Moreover, the feedback provided by smart contract developers is crucial in selecting the most useful tools.

As for the search strategy, we followed a protocol presented by Kitchenham for systemic reviews \cite{KITCHENHAM20097}. This protocol consists of three phases: (a) Define a search string; (b) Use the string in search engines; and (c) Select literature based on predefined inclusion and exclusion criteria. 

\begin{figure}
    \centering
    \includegraphics[width=1\linewidth]{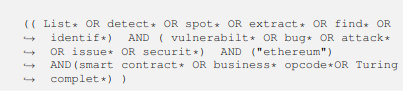}
    \caption{ Used search string }
\label{search-string}
\end{figure}

We defined a search string presented in Listing \ref{search-string} to collect the following resources: (a) Review articles about smart contract security; (b) Papers that discuss vulnerabilities in Ethereum smart contracts; (c) Papers that cover detection methodologies or mitigation schemes of smart contract vulnerabilities; and (d) Papers about tools that detect vulnerabilities in smart contacts. We have only included papers containing the keywords in the subject, title, or abstract. The term "opcode" refers to the basic instructions carried out by the blockchain's virtual machine, which regulates the logic and operations of the smart contract. Turing completeness, on the other hand, refers to the capacity of a blockchain or smart contract language to simulate any Turing machine. This means it can resolve any computational problem given enough time and resources. Both those terms are used to represent smart contracts.

\begin{figure}[h]
    \centering
    \includegraphics[width=0.48\textwidth]{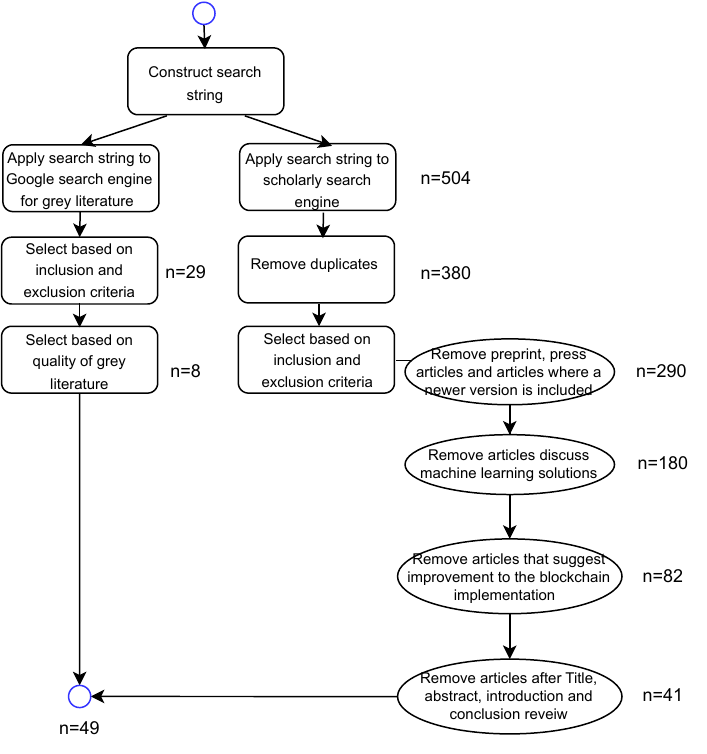}
    \caption{Articles identification, exclusion and inclusion methodology.}
    \label{fig:methodology}
\end{figure}

 Figure \ref{fig:methodology} illustrates the complete process of article identifications. 504  conference and journal articles were found using the specified search string. 41 articles are included based on the exclusion and inclusion criteria mentioned in table \ref{tab:criteria}. As for gray literature, we focused on blog posts found within the first eight pages of Google search results. The search strings include "Ethereum smart contract vulnerabilities and mitigation" and "Ethereum smart contract analysis tools." Out of the numerous blogs reviewed, eight posts met our established criteria.
To assist the quality of the gray literature, we have taken into consideration the following aspects:
\begin{itemize}
    \item The reputation of the publisher.
    \item The author's expertise in smart contract security, such as job position.
    \item The content is clearly stated with supported arguments.
\end{itemize}

\begin{table}

\setlength{\tabcolsep}{3pt}
\begin{tabular}{|p{35pt}|p{170pt}|}
\hline
Criteria & Description \\
\hline
    Inclusion Criteria &
  - Studies related to vulnerabilities in Ethereum: Vulnerabilities explication and mitigation techniques.
    \newline
   - Studies related to the security improvement of Ethereum smart contracts.
    \newline
   - Studies related to vulnerability detection tools in Ethereum smart contracts.
    \newline
    - Studies that introduce open-source tools to detect the vulnerabilities of solidity smart contracts.
     \\ 
  \hline
Exclusion Criteria  &
 - Non-English papers.
   \newline
  - Results past the first eight Google pages, as we have noticed that after the eight pages, the results are not relevant.
   \newline
   - Data sets, tweets, presentations.
    \newline
  - Tools that use machine learning, such as comparing machine learning tools, need different criteria, such as the training models, accuracy, etc.
   \newline
   - Studies that are based on the improvement of blockchain infrastructure rather than smart contract programming
   \newline
   - Study that is an older version of another paper that is previously been considered.
    \\ 
  \hline
\end{tabular}
\caption{Exclusion/inclusion criteria}
  \label{tab:criteria}%
\end{table}

\section{Vulnerabilities}
\label{vul}
In this section, we present some of the common vulnerabilities of Ethereum smart contracts alongside real-world attacks and prevention mechanisms.

\subsection{Reentrancy}
\subsubsection{Description}
Reentrancy is one of the most critical vulnerabilities to address when implementing smart contracts, also known as a recursive call attack. A contract calling another contract, with external calls, will cause the stoppage of the calling contract's execution and memory state until the called function returns a response. External calls are exposed in contract interfaces; hackers can use them to invoke a function within the contract numerous times, causing the contract to perform unanticipated tasks. This vulnerability occurs in a solidity smart contract performing critical tasks (e.g., token transfer) before resolving the effects that should have been addressed (e.g., balance update).

\subsubsection{Implications on NFT Fractionalization}
Reentrancy vulnerabilities have significantly impacted decentralized finance (DeFi) protocols, illustrating potential risks for fractionalized NFT platforms. For instance, the dForce DeFi Protocol Hack \cite{DFORCE} in February 2023, where an attacker exploited a reentrancy vulnerability in the Curve Finance vault on the Arbitrum and Optimism blockchains, part of the dForce protocol, led to the theft of approximately \$3.6 million in assets. This attack underscores the danger reentrancy poses to smart contracts handling complex financial transactions, serving as a warning for platforms dealing with fractionalized NFTs. Furthermore, the CREAM Finance Hack \cite{CREAM} in August 2021 and the Siren Protocol Hack \cite{SIREN} in September 2021, with losses of \$18.8 million and \$3.5 million, respectively, further exemplify the ongoing threat of reentrancy attacks to the blockchain and DeFi sectors.

Reentrancy attacks can distort market dynamics by unfairly redistributing assets, leading to market manipulation and loss of liquidity. For fractionalized NFTs, where valuation depends on the underlying asset's perceived value and the platform's integrity, such attacks can result in volatile price swings and diminished market confidence.

\subsubsection{Protection Measures} 
It is recommended to use \emph{send()} and \emph{transfer()}  methods instead of the general \emph{call()} method to transfer money. They are deemed safer because they have a gas limit of 2,300 gas. At the same time, the method \emph{call()} does not have a gas limit and forwards the remaining gas to the target address. The gas limit prohibits the target contract from making costly external function calls.
 The\emph{ checks-effects-interactions} pattern  \cite{sec-pattern} is the most reliable approach to prevent reentrancy attacks. This pattern defines how the code of a function should be organized to minimize undesirable side effects and execution behaviors. The programmer needs to include all checks, which usually consist of \emph{assert()} and \emph{need()} modifiers. If these checks pass, the function will resolve all of the effects of the contract ( e.g., balance update). The function should only communicate with other contracts once all state changes have been updated. Even if an attacker performs a recursive call to the initial function, the user cannot exploit the state of the contract since external functions are called last.
 
 Another protection mechanism against reentrancy attacks is the use of mutex. A mutex locks down the contract state. 
Mutex allows only one execution of a critical code section at a time using a lock mechanism. Only the lock's owner can modify it. When an attacker tries to perform a reentrancy attack, the previous call will lock the state, and the balance can not be updated. Mutex must be treated with care to ensure the lock can be released. \emph{ReentrancyGuard} is OpenZeppelin's own mutex implementation \cite{ReentrancyGuard}. This library offers a \emph{nonReentrant} modifier that secures an external function with a mutex.

\subsubsection{Example}
\begin{figure}
    \centering
    \includegraphics[width=\linewidth]{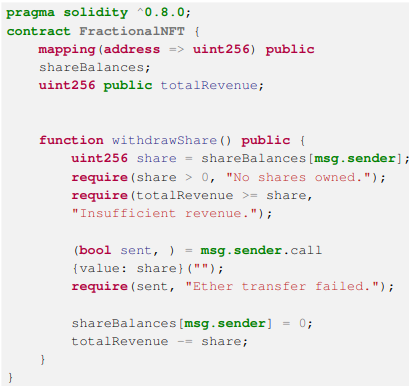}
    \caption{Simplified contract with reentrancy vulnerability }
\label{sm-Reentrancy}
\end{figure}

 Listing \ref{sm-Reentrancy} illustrates a simplified smart contract, FractionalNFT, designed to manage fractional shares of an NFT's revenue. This contract allows owners of fractional shares to withdraw their proportion of sales revenue when the NFT is sold. The contract contains a critical flaw in its \emph{withdrawShare} function, which makes it susceptible to reentrancy attacks. In this contract, the \emph{withdrawShare} function fails to adhere to the checks-effects-interactions pattern, updating the shareholder's balance \textit{after} transferring funds. This ordering allows for the potential reentrancy, where a malicious actor could recursively call \emph{withdrawShare} within a fallback function, draining the contract's funds beyond their rightful share.

\begin{figure}
    \centering
    \includegraphics[width=1\linewidth]{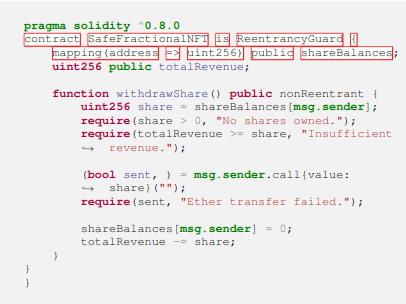}
    \caption{Simplified contract that mitigates reentrancy vulnerability }
\label{sm-ree-sol}
\end{figure}

To address this vulnerability, the \emph{ReentrancyGuard} utility from the OpenZeppelin security library can be integrated to prevent recursive calls, as demonstrated in Listing \ref{sm-ree-sol}. By employing the \emph{nonReentrant} modifier provided by \emph{ReentrancyGuard}, the revised contract ensures that the \emph{withdrawShare} function cannot be re-entered while it is still executing, effectively mitigating the reentrancy vulnerability.

 \subsection{Front Running}
\subsubsection{Description}
Front-running, also known as transaction ordering dependency, occurs when the execution logic depends on the order of the submitted transactions. The miner \cite{9152675} determines the order of transactions in Ethereum. The transaction is visible to the network before being executed. The participants can exploit this visibility by sending transactions with a higher gas price to be included first.   
\subsubsection{Implications on NFT Fractionalization}
Front-running can have profound implications in the context of NFT fractionalization that might include:

1. \textbf{Manipulation of Share Prices:} Malicious actors could exploit transaction ordering to manipulate the market for fractional tokens, buying up tokens at lower prices before a significant transaction increases their value or, conversely, selling tokens to depress prices ahead of a significant sell order.

2. \textbf{Interference with Auction Mechanisms}: Many NFT platforms use auction mechanisms for selling fractional shares or entire NFTs. Front-runners could preemptively place bids to disrupt fair auction outcomes, affecting the final sale price of an NFT.

3. \textbf{Unfair Distribution of Revenue}: In scenarios where revenue from NFT sales is distributed among shareowners, front-runners could strategically insert transactions to claim a disproportionate share of the distributions, undermining the fairness of the process.

\subsubsection{Protection Measures} 
The best solution to protect against front-running vulnerability is to remove the advantage of transaction ordering from the application. A solution is to remove the importance of time. Another possible solution is to use a commit-reveal hash scheme where the participant submits the hash of the answer instead of the answer. The contract then stores the hash and the sender's address; the answer is revealed only after all the responses are submitted.
 
\subsubsection{Example}

\begin{figure}
    \centering
    \includegraphics[width=1\linewidth]{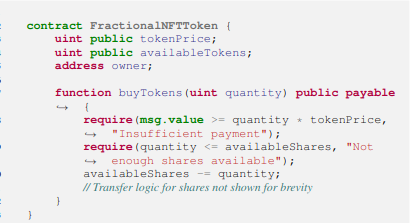}
    \caption{Simplified contract with Front Running vulnerability }
\label{sm-fr}
\end{figure}

Listing \ref{sm-fr}  illustrates a smart contract managing the sale of fractional tokens of an NFT, vulnerable to front-running. In this scenario, an attacker could observe pending purchase transactions and execute their purchase with a higher gas fee, securing shares at the current price before a large purchase increases their value, thereby disadvantaging legitimate buyers.

\begin{figure}
    \centering
    \includegraphics[width=1\linewidth]{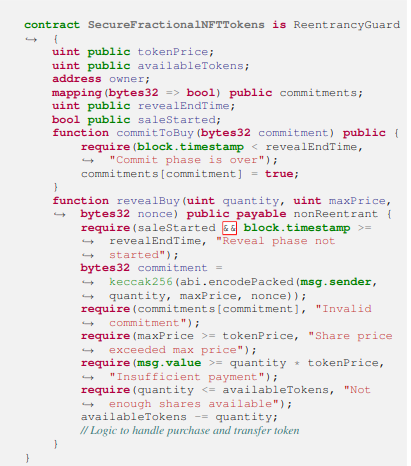}
    \caption{Simplified contract that mitigates Front Running vulnerability }
\label{sm-fr-sol}
\end{figure}

The contract can be revised to include a commit-reveal scheme to mitigate the vulnerability and ensure fair transaction processing. This modification requires buyers to commit to a purchase without initially revealing the exact quantity or price. This is followed by a reveal phase where the transaction details are disclosed as shown in Listing \ref{sm-fr-sol}.

\subsection{Arithmetic}
\subsubsection{Description} 

Solidity's integer types, such as \texttt{uint8}, \texttt{uint16}, and \texttt{uint256}, are constrained by fixed sizes, capable of representing numbers within specific ranges. Arithmetic operations that result in values outside these permissible ranges lead to integer overflow (exceeding the maximum value) or underflow (dropping below zero) \cite{smartcheck}. These vulnerabilities can cause smart contracts to behave unpredictably, potentially leading to significant security breaches or financial losses.

\subsubsection{Implications on NFT Fractionalization}
In the context of NFT fractionalization, arithmetic vulnerabilities pose unique challenges. For instance, when distributing sales revenue among shareholders, overflow or underflow errors can lead to incorrect allocation of funds. This affects the fairness and accuracy of fee distribution and exposes the platform to potential exploitation. Arithmetic vulnerabilities, particularly overflows and underflows, have had substantial consequences in the decentralized finance (DeFi) sector, highlighting the associated risks for platforms dealing with fractionalized NFTs. A prominent example includes the \emph{mintToken} function in the Coinstar (CSTR) Ethereum token's smart contract \cite{overflow}, which suffered from an integer overflow. This flaw permitted the contract owner to adjust any user's balance to any chosen value arbitrarily.
 Another example is the \emph{4chan gang group} experienced a substantial loss of \$2.3 million due to an underflow vulnerability in the ERC-20 token implementation of PoWH (Proof of Weak Hands), which allowed an attacker to exploit this flaw for financial gain  \cite{8668768}.

\subsubsection{Protection Measures} 
 Arithmetic operations should be appropriately implemented by checking the operators and operands before operating to avoid integer overflows and underflows. It is recommended to use the \emph{assert()} and \emph{require()} modifiers.
Using the library for arithmetic functions is also advisable, called \emph{SafeMath} by OpenZeppelin \cite{smartcheck}. Solidity version 0.8.0 has included this library, so the transaction will revert if an overflow/underflow occurs. A prevalent method for detecting this type of vulnerability involves taint analysis, a technique we will elaborate on in Section \ref{Detection}.

\subsubsection{Example} 

\begin{figure}
    \centering
    \includegraphics[width=1\linewidth]{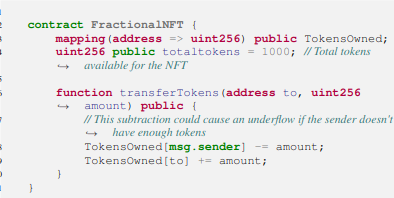}
    \caption{Simplified contract with Arithmetic vulnerability }
\label{sm-Arithmetic}
\end{figure}

Listing  \ref{sm-Arithmetic} showcases a smart contract managing fractional tokens of an NFT, vulnerable to underflow. If an attempt is made to transfer more tokens than available under Solidity versions before 0.8.0, this could lead to an underflow, setting the token balance to an incorrect high value and potentially enabling malicious token distribution.

\begin{figure}
    \centering
    \includegraphics[width=1\linewidth]{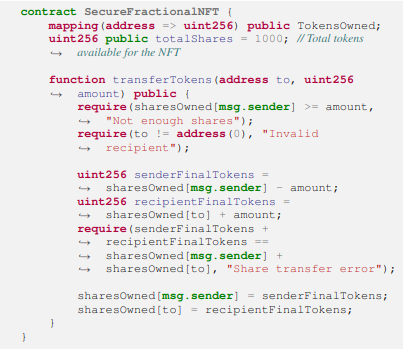}
    \caption{Simplified contract that mitigates arithmetic vulnerability}
\label{sm-Arithmetic-sol}
\end{figure}

Given Solidity 0.8.0's automatic checks for arithmetic operations, the compiler directly addresses the primary vulnerability of underflows and overflows. However, we can implement additional logic to further safeguard the integrity of fractional token calculations and transfers. An improved version of the smart contract is presented in Listing \ref{sm-Arithmetic-sol}.

\subsection{Mishandled Exceptions}
\subsubsection{Description}

Solidity provides two primary paradigms for interacting with external contracts: direct contract calls and low-level calls. Direct contract calls involve invoking functions directly on known contract interfaces, which inherently revert to failure, thereby throwing an exception. Conversely, low-level calls, executed via methods like \texttt{call()}, \texttt{delegatecall()}, and \texttt{callcode()}, return a boolean success flag instead of reverting on exceptions \cite{9667515}. These calls do not inherently revert transaction execution upon failure; instead, they return \texttt{false}, necessitating explicit checks of their return values to ensure the intended execution flow. Failure to adequately check the result of a low-level call can lead to unintended execution continuation, potentially compromising contract logic and security. This vulnerability was notably exploited in the King of Ether game, where the smart contract's failure to verify the result of a \texttt{send()} operation led to discrepancies in payments, resulting in users overpaying or underpaying \cite{kingeth}.

\subsubsection{Implications on NFT Fractionalization}
Mishandled exceptions, notably in operations involving the transfer of tokens or the distribution of revenues, pose significant risks to the platform's functionality. For example, if a smart contract designed to distribute sales revenue from an NFT among its token holders neglects to confirm the success of these transactions, it could lead to financial disparities. Such scenarios may arise when the contract employs low-level calls for fund transfers without verifying their execution success. Malicious entities might seize on these vulnerabilities, intentionally causing transactions to fail silently by making a contract reject transactions in its fallback function.

\subsubsection{Example} 
\begin{figure}
    \centering
    \includegraphics[width=1\linewidth]{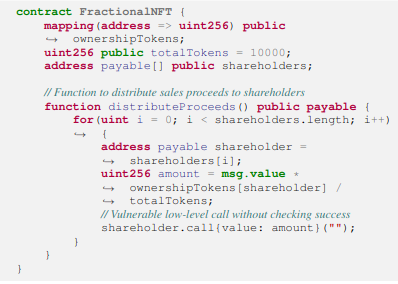}
    \caption{Simplified contract with mishandled exceptions}
\label{sm-mishandled}
\end{figure}

The example \ref{sm-mishandled} shows a contract that manages fractional ownership of an NFT. It includes a function to distribute proceeds from NFT sales to token holders. The contract incorrectly handles a low-level call when sending proceeds, posing a risk if the call fails.

\begin{figure}
    \centering
    \includegraphics[width=1\linewidth]{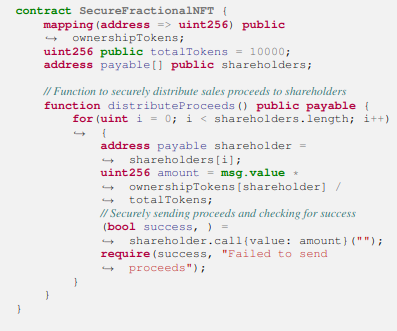}
   \caption{Simplified contract that mitigates mishandled exceptions}
\label{sm-mishandled-sol}
\end{figure}

In the revised contract \ref{sm-mishandled-sol}, we use a safer approach to distribute proceeds by checking the success of each payment and reverting the transaction if a payment fails.

\subsection{Code Injection via delegatecall}
\subsubsection{Description}
There is a unique method called a delegate call. The DELEGATECALL opcode is similar to a conventional message call, except that the code performed at the target address is executed in the context of the calling contract. The current address, storage, and balance refer to the calling contract. This dynamically enables a smart contract to load code from another smart contract at runtime. 
Calling into untrusted contracts via \emph{delegatecall()} is particularly risky since the code at the target smart contract has complete control over the caller's balance; thus, it can modify any of the caller's storage data.

\subsubsection{Implications on NFT Fractionalization} 
Malicious code executed through delegatecall might tamper with the contract's ownership or control mechanisms, enabling attackers to reroute assets or funds. Furthermore, attackers could change the revenue distribution logic, redirecting profits meant for rightful owners to unauthorized entities. 
In extreme cases, like the second Parity multi-sig attack where an attacker took over three main Parity wallets and stole \$31 million \cite{Parity}, these vulnerabilities could let attackers take over the NFT fractionalization contract entirely. They could push out the real owners and managers, potentially locking, freezing, or stealing assets.

\subsubsection{Protection Measures} 
 
When employing \emph{delegatecall()}, caution is paramount, especially when dealing with contracts not fully trusted. It's critical to avoid making delegatecall() to addresses derived from user inputs unless there's a rigorous verification process against a list of trusted contracts. This precaution helps ensure that only known, secure contracts can be interacted with, significantly reducing the risk of malicious interference.

Solidity's \emph{library} keyword facilitates the creation of library contracts designed to be stateless and immune to destruction. By defining a contract as a stateless library, it's guaranteed that the executing code cannot alter the storage data of the calling contract. This design principle is crucial for minimizing risks associated with storage context issues that \emph{delegatecall()} might introduce. Ensuring that delegatecall() is only used with contracts declared as libraries is a solid practice.

\subsubsection{Example} 

\begin{figure}
    \centering
    \includegraphics[width=1\linewidth]{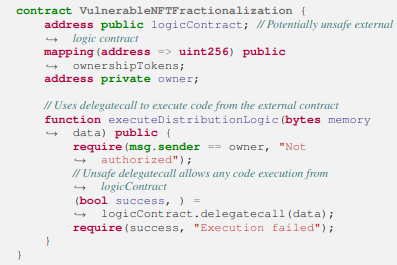}
  \caption{Simplified contract with Code Injection via delegatecall vulnerability }
\label{delegatecall}
\end{figure}

Listing \ref{delegatecall} demonstrates how an NFT fractionalization contract might be exposed to a code injection vulnerability through unsafe use of delegatecall. In this contract, \textit{executeDistributionLogic} can execute arbitrary logic via delegatecall based on the bytecode provided in the data. If \textit{logicContract} points to a malicious contract, it could lead to unintended alterations in its state, including token ownership manipulation.

\begin{figure}
    \centering
    \includegraphics[width=1\linewidth]{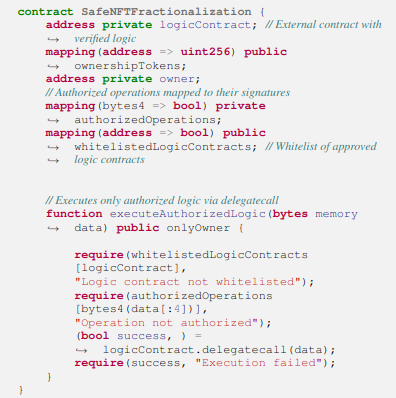}
  \caption{Simplified contract that mitigates Injection via delegatecall vulnerability }
\label{delegatecall-sol}
\end{figure}

To mitigate the risk of code injection via delegatecall, the contract should strictly control the update of the logicContract address and ensure that only verified, safe operations are executable. Also, we introduce a whitelist of approved logic contract addresses and require that any updates to \texttt{logicContract} come from this whitelist. Listing \ref{delegatecall-sol} illustrates these changes.

\subsection{Randomness Using Block Information}
\subsubsection{Description}
In blockchain-based applications, certain functionalities might require randomness—for instance, distributing a rare NFT fractionally among participants or deciding the winner of a unique NFT in a lottery; block information, such as block hash and block timestamp, can be used to achieve this. This information, however, may be anticipated and slightly modified by miners. When the block timestamp is used as the trigger condition to execute the transaction, it creates a vulnerable situation; dishonest miners can exploit the value of the block timestamp in an unethical manner. This vulnerability was exploited in GovernMental, a Ponzi scheme game \cite{Atzei}. The player who joined the round last and stayed for at least a minute was compensated according to the game rules. A miner who is also a player might change the timestamp to make it look like they were the last to join for more than a minute and, therefore, collect the reward.

\subsubsection{Implications on NFT Fractionalization} 
Randomness Using Block Information could impact Fractionalization solutions. For instance, if randomness derived from block attributes decides the allocation of rare NFT fractions or the winners of rewards, it could lead to outcomes unfairly attributed in favor of those with the ability to influence block information. Similarly, auctions for selling fractionalized NFTs could be manipulated, allowing miners or others with insider advantages to affect the auction's outcome by adjusting the voting period, for example.

\subsubsection{Protection Measures}

 Adopting more reliable sources of randomness is essential to address the vulnerabilities associated with using block information to generate randomness in smart contracts. One solution is employing oracles \cite{20193307309694} or other external sources that provide verified randomness robust solution.

Alternatively, cryptographic commitment schemes \cite{10.5555/646765.704125}, exemplified by solutions like RANDOA \cite{randao}, represent another practical approach. These schemes involve participants committing to their inputs in a concealed manner, which are later revealed collectively to generate a random outcome. This method ensures no individual can influence the result based on other participants' commitments, fostering fairness and security.

 \subsubsection{Example} 
 
\begin{figure}
    \centering
    \includegraphics[width=1\linewidth]{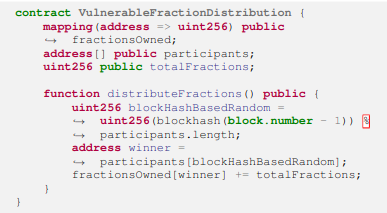}
    \caption{Simplified contract with Randomness Using Block Information vulnerability}
\label{Randomness}
\end{figure}

 Listing \ref{Randomness} shows an example of an NFT fractionalization platform that randomly assigns fractional shares of an NFT to participants based on block hash as a source of randomness. This contract is vulnerable because miners, or participants with mining capabilities, can potentially manipulate the block hash to influence the distribution outcome.

\begin{figure}
    \centering
    \includegraphics[width=1\linewidth]{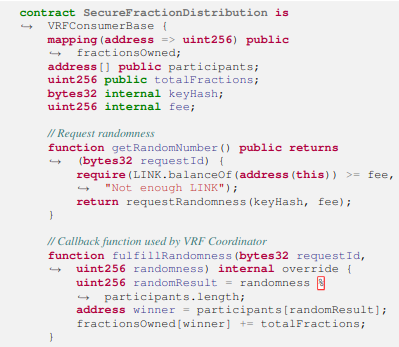}
   \caption{Simplified contract with Randomness Using Block Information vulnerability}
\label{Randomness-sol}
\end{figure}

To mitigate this vulnerability, the platform can utilize an external oracle to provide a source of verified randomness. This approach ensures that the randomness used for fraction distribution is not predictable or manipulable by miners. Contract illustrated in Listing \ref{Randomness-sol} inherits from VRFConsumerBase, provided by Chainlink \cite{VRF}, to securely request and receive verified random numbers. It uses the Chainlink VRF (Verifiable Random Function) to ensure that participants, including miners, cannot influence randomness in determining the distribution of NFT fractions.

\subsection{Other Vulnerabilities \& Summary}

Table \ref{tab:sumRefVul} presents the papers that include each vulnerability. We notice that the most mentioned vulnerability is reentrancy, whereas code injection is the least discussed.

\begin{table}[ht!]

\setlength{\tabcolsep}{3pt}
\begin{tabular}{|l|p{4.5em}|p{2em}|p{4em}|p{4.5em}|p{3em}|p{5em}|}
\hline
 & Reentrancy & Front Running & Arithmetic & Mishandled exceptions & Code Injection & Randomness \\
\hline
\cite{9050260} &  X & & & & &  \\
\cite{Vyper} & X & & X & & & X \\
\cite{ContractFuzzer} & X & & & X & X & X \\
\cite{ZEUS} & X & X & & & & X \\
\cite{10.1145/3183440.3183495} & X & & & & & \\
\cite{LOPEZVIVAR2021119} & X & & X & X & & X \\
\cite{Atzei} & X & & X & X & X & X \\
\cite{8726833} & X & X & & X & & X \\
\cite{9347617} & X & & & X & X & X \\
\cite{10.1145/3282373.3282419} & X & X & & X & & X \\
\cite{oyente} & X & X & & X & & X \\
\cite{EtherSolve} & X & & & & & \\
\cite{slither} & X & & & & & \\
\cite{smartcheck} & X & X & X & & & \\
\cite{securify} & X & X & & X & & \\
\cite{pattern} & & X & & & & X \\
\cite{Osiris} & & & X & & & \\
\cite{MadMax} & & & X & & & \\
\cite{10.1145/3385412.3385990} & & & & & X & \\
\hline    
  
\end{tabular}
\caption{\label{tab:sumRefVul} Coverage of vulnerabilities by each reference}
\end{table}

The Smart Contract Weakness Classification Registry (SWC) \cite{swcregistry} is a community database of known smart contract vulnerabilities; it is often up to date. Each issue includes a description, code samples, and protection measures. Currently, the registry contains 36 vulnerabilities. One has to check this registry often to stay up to speed on the newest threats.

In addition, Rusinek et al. \cite{SCSVS} proposed the Smart Contract Security Verification Standard (SCSVS). SCSVS is a 14-point checklist designed to standardize smart contract security for programmers, designers, security auditors, and vendors. By offering recommendations at every level of the smart contract development cycle, SCSVS helps avoid the most known security concerns and vulnerabilities.

\section{Detection methods}
\label{Detection}
In this section, we present the most common approaches for detecting smart contract vulnerabilities. These include static analysis, dynamic analysis, and formal verification. We will briefly describe each approach and identify its benefits and limitations.
\subsection{Static Analysis}
Static code analysis is a technique of debugging that involves reviewing source code before running it; this technique is also known as white-box testing \cite{Wogerer05asurvey}. It is accomplished by comparing a set of codes to predefined coding rules. There are several techniques to examine static source code; they can be incorporated into a single solution. Compiler technologies are frequently used to develop these techniques, such as Taint Analysis, and Data Flow Analysis \cite{owaspStatic}. 
This section presents some static analysis techniques, namely Control Flow Graph, Taint Analysis, and Symbolic Analysis.

\subsubsection{Control Flow Graph}
\begin{figure}[h]
    \centering
    \includegraphics[width=0.4\textwidth]{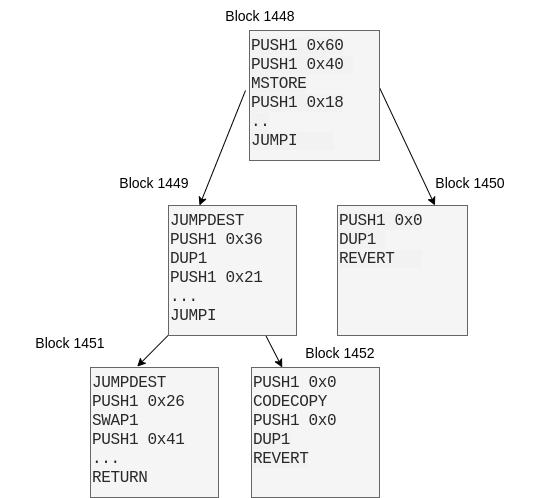}
    \caption{Example Control Flow Graph}
    \label{fig:cfg-ex}
\end{figure}

\hfill\\
The Control Flow Graph (CFG) is a directed graph describing the control flow. The flow of execution of source code A graph node denotes a basic block with no jumps; directed edges represent jumps from one block to another. If a node only has an exit edge, it is referred to as an \emph{entry}. And if it has only an entry edge, it is referred to as an \emph{exit} block. In Figure \ref{fig:cfg-ex}, Block 1448 is an entry block, and blocks 1451 and 1452 are exit blocks.

\begin{figure}[h]
    \centering
    \includegraphics[width=0.48\textwidth]{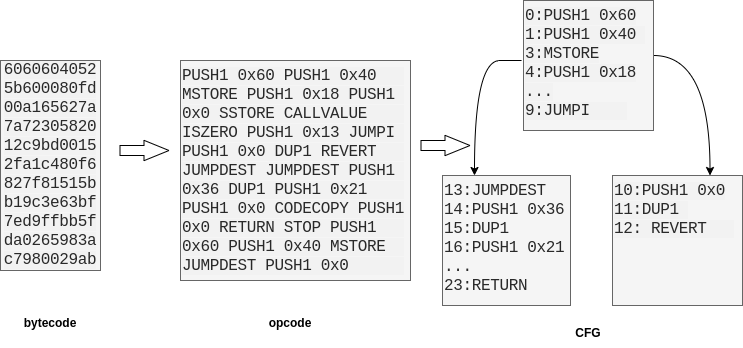}
     \caption{\label{fig:cfg}Process of generating CFG \cite{wood2014Ethereum}}
\end{figure}

Figure \ref{fig:cfg} shows the standard process to create CFG in the context of Ethereum. The first step is \textbf{Parsing bytecode}. Typically, the algorithm will extract the compiler version from the metadata; then, it will parse the remaining bytecode (without the metadata) into opcodes \cite{opcodes}. The parsing stage is straightforward because each bytecode's two characters represent one opcode (0x04 represents DIV). The complete list of opcodes can be found in the Ethereum yellow paper \cite{wood2014Ethereum}. The second step is \textbf{Identification of basic blocks}. A basic block is a set of opcodes that run in succession between a jump target and a jump instruction, with no other instructions interrupting the control flow. To determine the basic blocks, we need to identify the opcodes that alter the execution of the control flow; some examples are presented in Table \ref{tab:opexecution}. The JUMPDEST instruction starts a new basic block, whereas the other opcodes terminate blocks. Each block is identified by an offset representing its first opcode's position in the bytecode. The last step will be computing the edge by checking the destination offset; this step is not trivial as the destination is not an opcode parameter. In the literature, there are two types of jumps: (a) A \emph{pushed jump} is a JUMP followed by a PUSH opcode, allowing the target offset to be determined simply by glancing at the data in the previous PUSH opcode; and (b) \emph{Orphan jump} is not followed by a PUSH; indeed, its destination is not computed directly \cite{DBLP:journals/corr/abs-2103-09113}. The computation of orphan jumps depends on each implementation and is critical for the completeness of the graph.

\begin{table}

\setlength{\tabcolsep}{3pt}
\begin{tabular}{|p{33pt}|p{56pt}|p{126pt}|}
\hline
Operation& 
Opcode& 
Description \\
\hline
       Flow &  JUMP & Alters the program counter  \\
        ~ &  ~ & Alters the program counter  \\
        Operations  &  JUMPI & Conditionally alters the program counter  \\

~ & JUMPDEST &  Marks a valid destination for jumps \\
       \hline
  
           System & STOP & Halts execution   \\
         operations  &   REVERT &
Halts execution reverting state changes  \\
~ & ~ &  but returning data and remaining gas \\
~ &  RETURN &
 Halts execution returning output data \\
 ~ & INVALID & Designated invalid instruction \\
   ~ & SELFDESTRUCT &
Halts execution and registers account \\
~ & ~ & for later deletion \\
\hline    
  
\end{tabular}
\caption{\label{tab:opexecution}Opcodes that alter execution \cite{wood2014Ethereum}}
\end{table}

\subsubsection{Taint Analysis}
\hfill\\
Taint Analysis aims to detect input variables where data comes from an untrusted source that could be controlled by an attacker (e.g., environmental data or function parameters). These data inputs are \emph{tainted} by the taint analysis tool and then traced back to potentially sensitive functions, such as security checks or storage access, often known as \emph{sinks}. If the tainted variable flows into a sink, it is marked as a vulnerability \cite{Osiris}. 

In the context of Ethereum smart contracts, this technique is usually used after generating CFG. It exploits the opcode that an attacker may employ to insert data into the control flow; it can be divided into block, environmental, and flow information. Table \ref{tab:opinsertion} shows an example of an opcode that allows data insertion can be found in.

\begin{table}

\setlength{\tabcolsep}{3pt}
\begin{tabular}{|p{40pt}|p{61pt}|p{114pt}|}
\hline
Operation& 
Opcode& 
Description \\
\hline
 Block  & GASLIMIT & Gets the current block gas limit  \\
  information  & TIMESTAMP & Gets the current block timestamp \\
  
         \hline
      
           Environment & CALLER & Gets caller address  \\
         information  &  CALLDATALOAD &  Gets input data of current \\
         ~ & ~ & environment \\
         ~ & CALLDATACOPY & Copies input data in current  \\
        
         ~ & ~ & environment to memory \\
       \hline
     
           Flow & SLOAD & Loads word from storage  \\
         Operations  &  MLOAD & Loads word from memory \\

    \hline  
  
\end{tabular}
  \caption{ Opcodes that allow data insertion \cite{wood2014Ethereum}}
  \label{tab:opinsertion}
\end{table}

An example of taint analysis is to detect arithmetic bugs. The first step is to define the possible entrusted inputs (source), such as CALLDATALOAD and SLOAD. The affected memory, storage, or stack location is tagged. The next step is to check whether the flow of tagged input contains arithmetic operations such as ADD, MUL, and SUB with no catch mechanism (sink). For example, a catch mechanism is reverted in the case of DIV with 0. If the tagged input falls into a sink, it will be marked as a possible vulnerability. 

\subsubsection{Symbolic Analysis}
\hfill\\
Symbolic execution is a way of abstracting the execution of a program such that it can span numerous pathways through the code. The program is run with \emph{symbols} as inputs, and the outputs are expressed in the form of the symbolic inputs. Each symbolic path has a condition; it is a formula created by collecting constraints that must be satisfied by those inputs for the execution to continue on that path. If the condition is unsatisfiable, the path is infeasible; otherwise, the path is feasible \cite{10.1145/360248.360252}. 
Symbolic analysis is usually paired with CFGs. 
To find exploits, most tools employ the Satisfiability Modulo Theories (SMT) \cite{barrett2018satisfiability} solver to check whether the symbolic output is feasible or provide a counterexample if it is not.

\subsection{Dynamic Analysis}
"dynamic analysis" refers to observing code while executed in its original context. It is also known as black-box testing because it is performed without access to the source code. It works the same way as an attacker who feeds malicious code or unpredictable input to the appropriate functions of a program to look for vulnerabilities. The most common technique is Fuzzing.

\subsubsection{Fuzzing}
\hfill\\
Fuzzy testing, often known as fuzzing, is a kind of automated software testing that involves injecting incorrect, malformed, or unpredictable inputs into a system; the objective is to uncover software flaws and vulnerabilities. A fuzzing tool injects these inputs into the program and then watches for problems such as crashes or data leaks; it also can perform static analysis on execution traces. In the case of smart contracts, Wang et al. \cite{Wang2020} deploy the smart contract on a test network; then, they monitor the balance of the smart contract to identify basic misappropriation. This technique removes the necessity for particular patterns to determine a vulnerability.

\subsection{Formal Verification}
Formal verification is a technique that automates bug detection of a hardware or software system by comparing a formal system model to formal requirements or behavior specifications. Through mathematical analysis, formal verification can provide a high level of confidence.

To conduct a formal verification, we first need to provide specifications. A program specification is an unambiguous definition of the program's purpose and the scenarios that are allowed or not to be executed. Tools usually build deduction trees to verify a property where the root presents a Hoare triplet \cite{10.1145/3167084}. A Hoare triplet consists of: (a) \emph{Pre condition}: It is the initial state; (b) \emph{Instructions}: A series of transactions; and (c) \emph{Post condition}: It is the property to verify. To build deduction trees, theorem-proving algorithms are usually used, such as Coq \cite{Annenkov_2020} and Isabelle/HOL \cite{10.1145/3167084}.

There are a few contributions that propose languages to write formal specifications. Permenev et al. \cite{verx} proposed a specification language called VerX,  where the syntax is similar to Solidity; it supports temporal logic. The Ethereum community also proposed a language named \emph{Act} \cite{act}, which is a specification for Ethereum Virtual Machine (EVM) programs.

\subsection{Comparison}
The goal of using detection methods such as static, dynamic, and formal verification is to ensure that smart contracts are correctly executed in all states; this is to ensure that no vulnerabilities are produced and that specifications are respected. Table \ref{tab:sumRefMeth} shows the papers that discuss each methodology.

Static code analysis tools can potentially produce false negative results; vulnerabilities occur but are not reported. This could happen because the analysis tool doesn't allow for many test scenarios; it doesn't go into detail when looking at different states because it usually gives you a timeout. Various vulnerabilities may be false negatives in static analysis; however,  they can be detected correctly utilizing dynamic analysis tools. Thus, it is strongly recommended that both types of analysis be combined.

Dynamic analysis generates transactions with random inputs, which means that it observes random states depending on the inputs. The challenge for these techniques is to generate enough transactions and carefully select inputs to achieve maximum coverage.

Formal verification can provide full coverage for the specification under consideration because it employs mathematical analysis. This guarantees the satisfaction of a property if verified; however, in general, we cannot cover everything because the verification can become quickly intractable. Thus, false negatives can still be present. In addition, formal verification requires a skilled programmer who can express a smart contract as a mathematical high-level specification while accounting for a specific low-level virtual machine. This operation takes a long time and requires a lot of resources. Audit firms usually make use of formal verification  \cite{consensys2}.

\begin{table}[h]

\setlength{\tabcolsep}{3pt}
\begin{tabular}{|p{76pt}|p{139pt}|}
\hline
Vulnerabilities& 
References \\
\hline
       Static analysis & \cite{ZEUS}  \cite{oyente}  \cite{Osiris} \cite{securify} \cite{Ethainter} \cite{slither} \cite{KEVM} \cite{smartcheck} \cite{MadMax} \cite{verx} \cite{EthVer} \\
       \hline
     Dynamic analysis & \cite{Maian} \cite{echidna} \cite{ContractFuzzer} \cite{CONTRACTLARVA} \cite{Harvey} \cite{Liu2018ReGuardFR} \cite{manticore} \\
       \hline
    Formal verification & \cite{inbook} \cite{KEVM} \cite{10.1145/3167084} \cite{ZEUS} \cite{EthVer}\\
     
\hline    
  
\end{tabular}
\caption{\label{tab:sumRefMeth} Detection methods References}
\end{table}

\section{Vulnerability Detection Tools}
\label{Vulnerabilities}
This section will cover the most popular tools for detecting smart contract vulnerabilities. To investigate available tools, we first looked through academic literature and review articles (see Table \ref{tab:tools}). Then, we covered five of the most commonly used tools, as indicated by developer forums, and the number of forks and update frequency on GitHub.
\begin{table}[ht!]
 
\begin{tabular}{| p{10pt}| p{60pt} | p{85pt}| p{35pt} | } 

 \hline
    
 \# & Tool & Detection Technique & Last update \\ 
  \hline

   ~ & ~ & ~ & ~\\
 1 &Slither \cite{slither} \cite{slithergithub} & Static Analysis & May 2023
 \\ 
 ~ & ~ & ~ & ~ \\
  
  2 &MythX  \cite{Mythx}& static, dynamic \& Symbolic Execution & Not Public \\ 
   ~ & ~ & ~ & ~\\
  
  3 & Mythril \cite{mythril}  \cite{mythrilgithub}& Symbolic Execution &  May 2023 \\ 
   ~ & ~ & ~ & ~\\
  4 &   Echidna  \cite{echidna} \cite{echidnagithub}   & Fuzzer  & Apr 2023 \\ 
    ~ & ~ & ~ & ~\\ 
5 &   Manticore \cite{manticore} \cite{manticoregithub}  & Symbolic Execution & June 2022 \\ 
 ~ & ~ & ~ & ~\\
6 &   Securify  \cite{securify} \cite{securifygithub}  &  Static Analysis & Sep 2021 \\ 
 ~ & ~ & ~ & ~\\
7 &   KEVM  \cite{KEVM} \cite{KEVMgithub}  &  Static Analysis &  Apr 2022  \\ 
 ~ & ~ & ~ & ~\\
8 &  Smartcheck  \cite{smartcheck}  \cite{smartcheckgithub}    & Static Analysis   & Dec 2019  deprecated  \\  ~ & ~ & ~ & ~\\

9 &   MadMax  \cite{MadMax} \cite{MadMaxgithub}   & static 
analysis  &Jun 2021 \\ 
 ~ & ~ & ~ & ~\\
10 &   Vertigo  \cite{vertigo} \cite{vertigosgithub}  & Mutation Testing  & Feb 2021 \\ 
 ~ & ~ & ~ & ~\\
11 & EtherSolve \cite{EtherSolve} \cite{EtherSolvegithub}& CFG Extraction & Nov 2021 \\ 
 ~ & ~ & ~ & ~\\
12 &   Octopus  \cite{octopusgithub}    & Symbolic Execution  & Nov 2020 \\ 
 ~ & ~ & ~ & ~\\
 13 &  Oyente \cite{oyente} \cite{oyentegithub} & Symbolic Execution & Nov 2020 deprecated \\ 
 ~ & ~ & ~& ~ \\

14 &   ERC20 Verifier  \cite{erc20-verifier}   &  Verify ERC20 Compatibility  & Nov 2019 \\
 ~ & ~ & ~ & ~\\
15 &   Solgraph  \cite{solgraphgithub}   & CFG Extraction  & Jan 2019 \\ 
 ~ & ~ & ~ & ~\\
16 &   Osiris  \cite{Osiris} \cite{Osirisgithub}  &  Symbolic Execution  & Sep  2018\\

  \hline
\end{tabular}
\caption{Ethereum Smart Contract Vulnerability Detection Tools}
  \label{tab:tools}%
\end{table}

\subsection{Oyente}
\subsubsection{Description}

 Luu et al. propose Oyente \cite{oyente}, the first symbolic execution tool for Ethereum smart contracts. Oyente was the basis of several tools developed later \cite{Maian} \cite{Osiris}. It has four components: (a) \emph{CFGBuilder}: It constructs CFG of the smart contract; (b) \emph{Explorer}: It takes as input the Ethereum state. It has a loop that runs a state and then executes an instruction on the output of that state. The loop continues until no state remains or a timeout is reached. The output is a symbolic trace. \emph{Explorer} determines the infeasible trace by querying the Z3 SMT solver \cite{z3}; (c) \emph{CoreAnalysis}:  It targets predefined vulnerabilities by looking for patterns on the symbolic trace; (d) \emph{Validator}: It queries Z3 solver with traces flagged as vulnerable to reduce false positives cases.

Oyente detects seven types of vulnerabilities: Re-entrancy, Integer overflow/underflow, Transaction order dependence, Timestamp dependence, Callstack Depth, EVM Code Coverage, and Parity Multisig bugs.

\subsubsection{Execution Example}

 For evaluating Oyente's capabilities in the context of smart contracts dealing with token sales and potentially fractionalized assets, we utilize the \textit{TokenSaleChallenge.sol} smart contract, sourced from the Capture the Ether challenge series \cite{tokensalechallenge}. The contract embodies a token sale mechanism where tokens are bought and sold at a fixed price, presenting an analogous scenario to NFT fractionalization where individual tokens could represent shares of a fractionalized NFT.

\begin{figure}[ht!]
    \centering
    \includegraphics[width=0.48\textwidth]{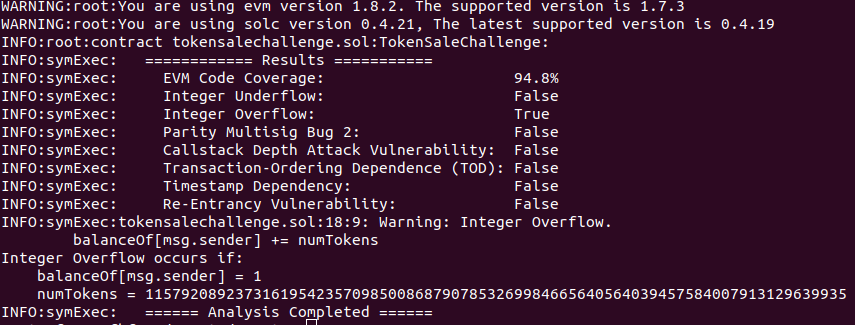}
    \caption{Oyente Output}
    \label{fig:Oyente}
\end{figure}

The primary aim is to assess Oyente's effectiveness in detecting vulnerabilities that could impact contracts handling fractionalized NFTs. The results, depicted in Figure \ref{fig:Oyente}, highlight an Integer Overflow vulnerability and the execution trace where this issue is detected. 

\subsubsection{Pros and Cons}

Oyente's deployment via a Docker image significantly simplifies its setup process, making it accessible for users with varying expertise in blockchain technology. This ease of use facilitates quick integration into development workflows, allowing for the immediate analysis of Ethereum smart contracts. Despite its user-friendly setup, Oyente's compatibility is somewhat limited; it supports Ethereum compiler versions only up to 0.4.17. This restriction may hinder its applicability to smart contracts compiled with newer versions of the Solidity compiler, potentially limiting its utility in analyzing the latest smart contract developments.

The output provided by Oyente is clear and concise, enabling developers and auditors to interpret vulnerability reports easily. Moreover, Oyente has played a pivotal role in evolving smart contract analysis tools. It has laid the groundwork for subsequent innovations, including Maian \cite{main} and Osiris \cite{Osiris}. These tools have built upon Oyente's foundational techniques, such as constructing Control Flow Graphs (CFGs), to offer enhanced accuracy in detecting vulnerabilities and orphan paths within smart contracts.

\subsection{Slither}
\subsubsection{Description}

 In their work, Feist et al. \cite{slither} unveiled a static analysis instrument known as Slither. This tool is aimed at spotting weaknesses, enhancing code efficiency, and augmenting the understanding of code. The procedure it employs is in multiple stages: (a) \emph{Data Retrieval}: Initially, it creates the Abstract Syntax Tree of the contract's source code using the Solidity compiler, yielding some information like the Control Flow Graph and contract inheritance; (b) \emph{Conversion into SlitherIR}: The source code of the smart contract is then converted into an internal representation language referred to as SlithIR; and (c) \emph{Examination of the Code}: This phase involves determining the variables being read and written as well as their types. Furthermore, it detects unsecured functions where potentially harmful addresses can execute high-level operations. During this \emph{Examination of the Code} stage, it also calculates data dependency and marks variables that rely on user input as tainted.

Slither is proficient in uncovering 70 bug varieties, encompassing self-destructive smart contracts, code injection via delegatecall, frozen ether, and Reentrancy vulnerabilities \cite{slithervul}.  

\subsubsection{Execution Example}

\begin{figure}
    \centering
    \includegraphics[width=1\linewidth]{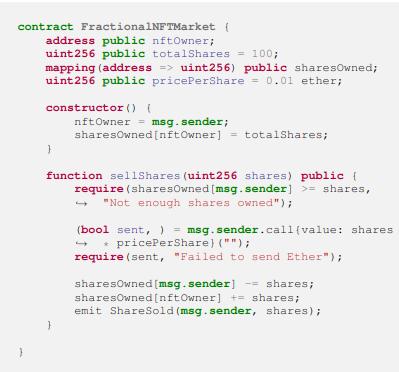}
  \caption{FractionalNFTMarket Smart Contract}
\label{sm-fractionalnftmarket}
\end{figure}

\begin{figure}[ht!]
    \centering
    \includegraphics[width=0.48\textwidth]{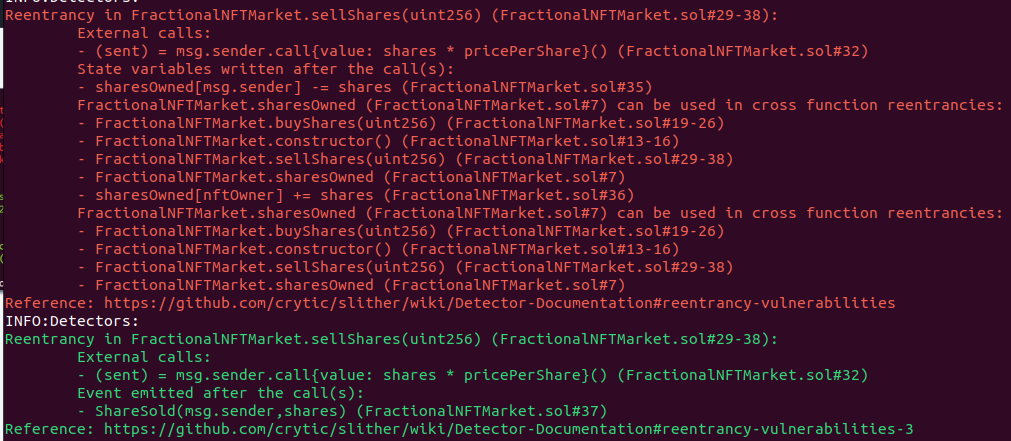}
    \caption{Slither Analysis Output for FractionalNFTMarket Contract}
    \label{fig:Slither}
\end{figure}

Using Slither to identify potential security vulnerabilities, we analyzed the \textit{FractionalNFTMarket} smart contract in Listing \ref{sm-fractionalnftmarket}. The analysis results are presented in Figure \ref{fig:Slither}. Slither has successfully detected a reentrancy vulnerability within the \textit{sellShares} function, which could allow an attacker to exploit the contract by recursively calling the function to drain its funds. In addition to detecting this critical vulnerability, Slither offered recommendations for code improvements and provided a comprehensive set of information, including a human-readable summary, an inheritance graph, and the Control Flow Graph (CFG) of each function.

Slither also provides a list of helpful information, including a human-readable summary of the scanned contract, inheritance graph, and CFG of each function; the complete list can be found in the GitHub repository \cite{slithergithub}.

\subsubsection{Pros and Cons}

Slither, an open-source static analysis tool, is recognized for its efficiency in auditing smart contracts. It is user-friendly, offering straightforward installation options via Docker or Python package managers. With the capability to detect approximately 70 types of vulnerabilities, Slither facilitates improved code understanding through visual aids like contract graphs. Moreover, it examines smart contract compliance with established Ethereum Request for Comments (ERC) standards, including ERC-20 and ERC-777 \cite{slConformance}.

One limitation of Slither is its tendency to report false positives, which can complicate the auditing process by requiring additional validation to review the findings. Crytic \cite{cryticio}, a premium service, can be utilized for a more comprehensive analysis. Crytic extends Slither's capabilities by identifying an additional 50 types of faults that Slither might miss. Furthermore, Crytic offers integration with GitHub, allowing for automated testing on pull requests, thereby enhancing the development workflow and ensuring continuous contract integrity.

\subsection{Mythril}
\subsubsection{Description}

 Slithe Durieux et al. \cite{mythril} developed a tool known as Mythril to identify potential weaknesses in smart contracts by utilizing symbolic execution. Mythril, a Python-based command-line tool, allows for interactive inspection of smart contracts. It produces a Control Flow Graph (CFG) and performs symbolic execution of EVM bytecode to restrict the search area. The tool can generate concrete values to exploit any detected vulnerabilities. To evaluate the feasibility of different paths, Mythril uses the Z3 SMT solver \cite{z3}.

Mythril can discover fourteen distinct types of vulnerabilities, including Delegate Call To Untrusted Contracts, Dependence on Predictable Variables, Deprecated Opcodes, Ether Thief, Exceptions, External Calls, Integer Over/Underflow, Multiple Sends, Self-Destruction, State Change External Calls, Unchecked Return Values, User Supplied Assertions, and Arbitrary Storage Write and Arbitrary Jump \cite{mythrilModules}.

\subsubsection{Execution Example}

\begin{figure}[ht!]
    \centering
    \includegraphics[width=0.48\textwidth]{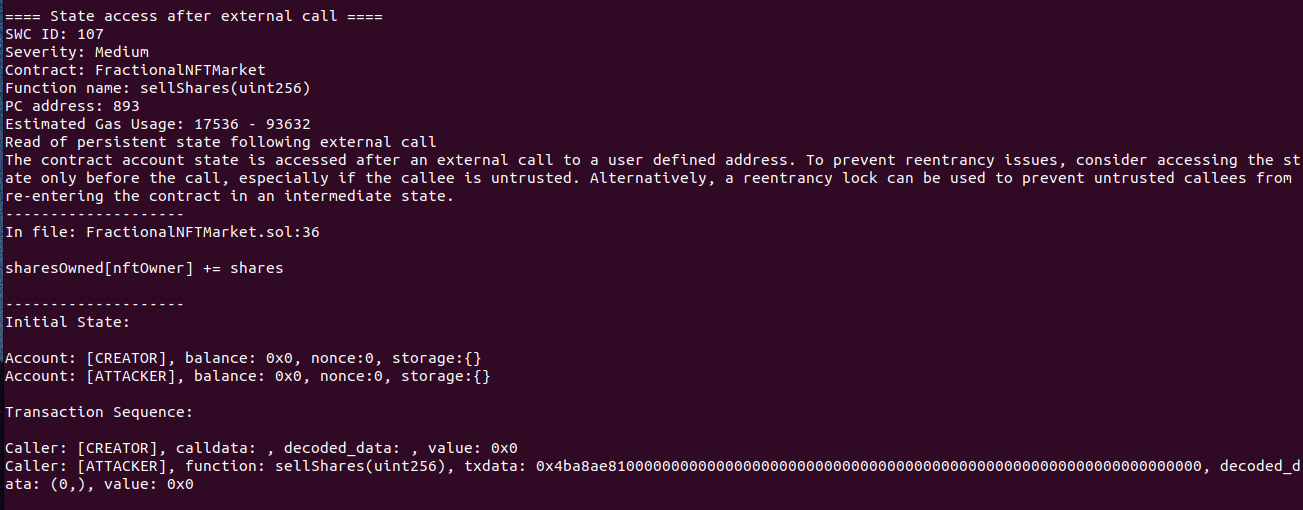}
    \caption{Mythril Output}
    \label{fig:Mythril}
\end{figure}

The \textit{FractionalNFTMarket.sol} smart contract, presented in Listing \ref{sm-fractionalnftmarket}, has been analyzed using Mythril to uncover potential security issues. The resulting output is depicted in Figure \ref{fig:Mythril}. Mythril has identified a reentrancy vulnerability within the contract’s `sellShares` function. The output provides detailed information, including the state of the contract, when the vulnerability can be triggered, and the specific transaction sequence that leads to the issue. Unlike Slither, Mythril offers the identification of vulnerabilities and the concrete inputs that cause them, enhancing the developer's understanding of the contract's weaknesses. Nevertheless, Mythril's output does not extend to include recommendations for remediation of the detected issues.

\subsubsection{Pros and Cons}
~ \\ Mythril stands out for its accessibility, offering straightforward installation options via Docker or Python package manager, streamlining the setup process. It allows for granular analysis, with the flexibility to test individual functions through custom Python scripts. This targeted testing can be particularly advantageous for developers seeking to debug specific aspects of their smart contract code.

However, Mythril's thorough approach to smart contract analysis can be computationally intensive, often resulting in longer analysis times than tools like Slither. The resource-heavy nature of Mythril's symbolic execution process may not be the most efficient choice for rapid iteration during development.
For those seeking a more advanced feature set and cloud-based capabilities, Mythril's commercial counterpart, MythX \cite{Mythx}, offers enhanced analysis power and can be seamlessly integrated into continuous integration/continuous deployment (CI/CD) pipelines \cite{mythrici}.

\subsection{Manticore}
\subsubsection{Description}

Mossberg et al. \cite{manticore} implemented a tool named Manticore to detect vulnerabilities in smart contracts. The primary aim of Manticore is to track inputs that kill a program, record instruction-level implementation, and provide Python API access to its analysis engine. Manticore uses symbolic execution to locate distinct computation paths in EVM  bytecode. It discovers inputs that will stimulate these computation paths with the aid of the SMT solver Z3 \cite{z3}. It keeps track of the execution traces for each execution. Manticore converts Solidity code to bytecode for evaluation; then, it examines the traces for vulnerabilities, such as reentrancy and reachable self-destruct operations, reporting them in the source code context. This tool is developed by \emph{TrailOfBits} \cite{trailofbitsm}. 

Mythril detects ten types of vulnerabilities, namely, delegatecall, overflow, reentrancy, Use of potentially unsafe/manipulable instructions, Reachable external, Reachable self-destruct instructions, Uninitialized memory usage, Uninitialized storage usage, Enable INVALID instruction detection, and 	Unused internal transaction return values \cite{manticoreModules}. 

\subsubsection{Execution Example}

For our examination of Manticore, we used the \emph{TokenSaleChallenge.sol} \cite{tokensalechallenge}, the same one used to test Oyente. This contract represents a simplified model for trading tokens, which can be seen as stand-ins for fractional tokens of an NFT. 

\begin{figure}[ht!]
    \centering
    \includegraphics[width=0.48\textwidth]{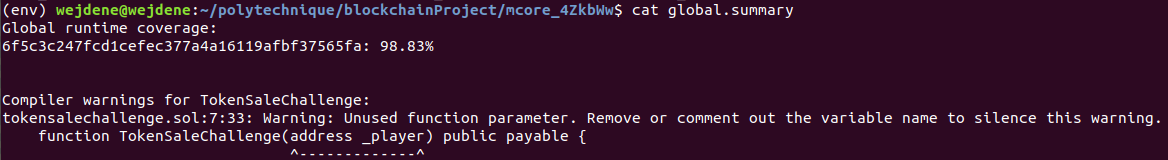}
    \caption{Manticore Output}
    \label{fig:manticore}
\end{figure}

The tool is extremely slow as it takes more than one hour to perform the test; it takes less than a minute for the other tools using the same smart contract and machine. The analysis output is a folder containing a summary report, traces, and analysis for each test case. Manticore generated 51 test cases but failed to detect the smart contact's reentrancy vulnerability, as shown in Figure \ref{fig:manticore}. 

\subsubsection{Pros and Cons}

Manticore offers a diverse set of command-line tools along with scriptable Python APIs, suitable for a variety of use cases and allowing for in-depth analysis procedures. It covers a broad spectrum of known vulnerabilities, making it a valuable asset for developers seeking to secure their smart contracts.

However, the detailed nature of Manticore's analysis comes with a trade-off in terms of performance. The tool can take significantly longer to complete its analysis than others, sometimes leading to timeouts. This is often due to its comprehensive symbolic execution approach, which, while powerful, is also resource-intensive. Additionally, Manticore requires considerable memory, which could be a limiting factor for users with constrained hardware resources.

\subsection{Echidna}
\subsubsection{Description}
Echidna is a specialized property-based fuzzing tool for smart contracts implemented by Grieco at al. \cite{echidna} that takes inspiration from QuickCheck \cite{QuickCheck}. Echidna seeks to violate user-defined invariants that represent potential contract errors rather than just finding crashes \cite{Echidnablog}. The tester writes these invariants, and it is called \emph{echidna property}: A particular Solidity function with no arguments that returns "true" on success and has a name that begins with "echidna." Echidna reports any transactions that result in these properties returning false or error, essentially disclosing contract bugs \cite{EchidnaTest}.
Echidna employs various strategies to produce test inputs, including iterative feedback, structural constraints, and known input variation.

\subsubsection{Execution Example}

\begin{figure}
    \centering
    \includegraphics[width=1\linewidth]{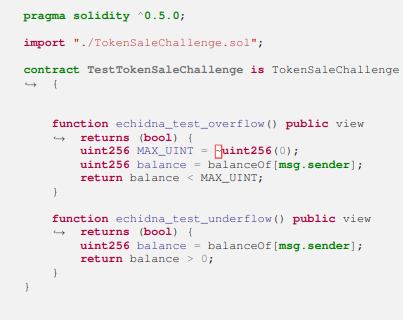}
   \caption{TestTokenSaleChallenge Smart Contract}
\label{TestToken}
\end{figure}

To test Echidna, we have adapted the \emph{tokensalechallenge.sol} smart contract \cite{tokensalechallenge}. We created a derived contract 
\emph{TestTokenSaleChallenge} that extends the original \emph{TestTokenSaleChallenge} contract shown in Listing \ref{TestToken}. The test properties in the \emph{TestTokenSaleChallenge} are designed to check for arithmetic errors that can occur in the logic used for NFT fractionalization.

The \emph{echidna\_test\_overflow} function aims to validate that token balances do not exceed the maximum uint256 value, a critical check to prevent overflow in token allocations, which could mirror similar concerns in the distribution of fractional NFT shares. Conversely, \emph{echidna\_test\_underflow} ensures that the balance never drops below zero.

 By defining these test functions, Echidna will attempt to generate test inputs that trigger these conditions to ensure the contract behaves as expected.

\begin{figure}[ht!]
    \centering
    \includegraphics[width=0.48\textwidth]{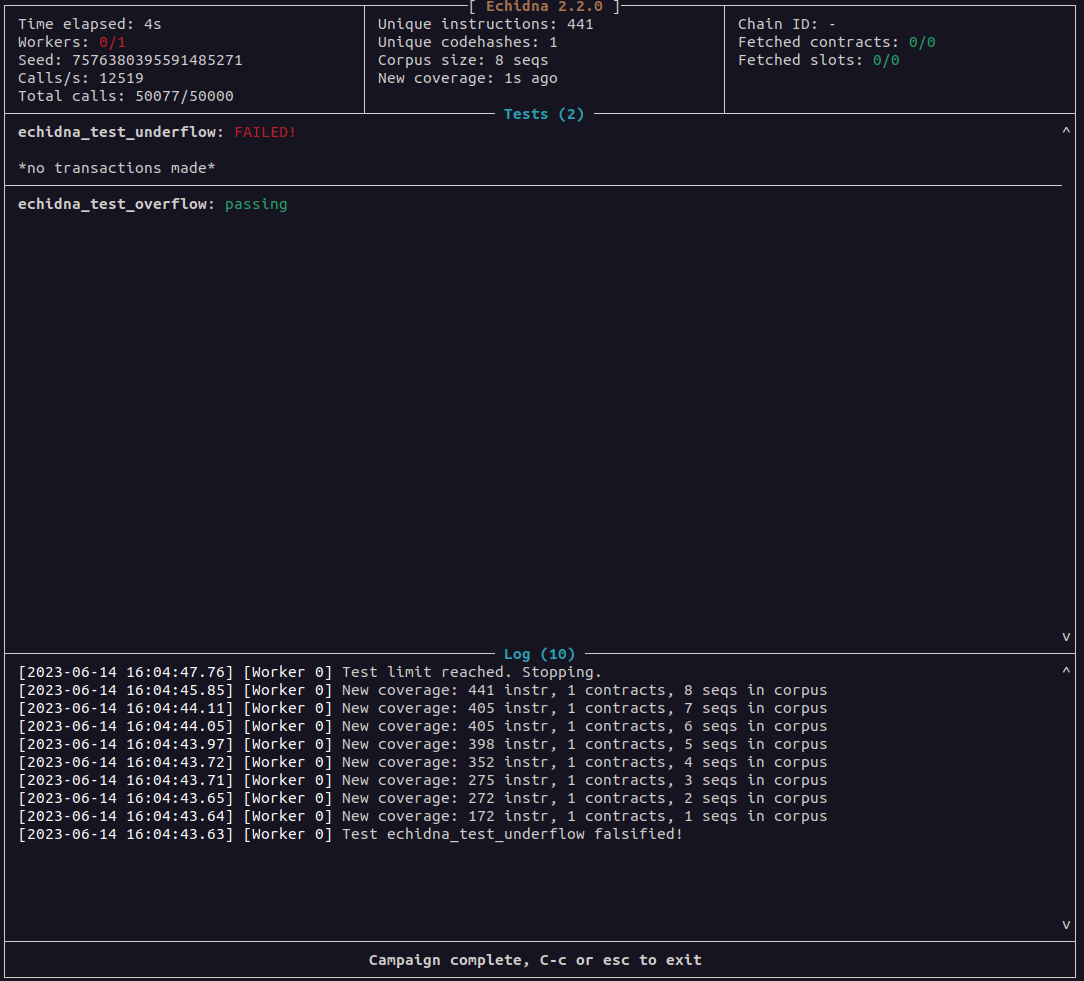}
    \caption{Echidna Output}
    \label{fig:echidna}
\end{figure}

Echidna generates its results directly to the console, as figure \ref{fig:echidna} shows, distinguishing between successfully validated and failed properties. Echidna offers a counterexample for each property that didn't meet the validation criteria, which is essentially a step-by-step breakdown of how the property failed under specific conditions.
In our example, the underflow test did not pass, even without initiating a transaction. This was because the msg.sender was left empty, indicating that the system can encounter failures in scenarios where sender information is not provided.
However, Echidna didn't detect an overflow issue, possibly due to a short timeout period, which prevented it from producing a sequence revealing this vulnerability as we did not change the default configuration.

\subsubsection{Pros and Cons}

One of Echidna's notable strengths is its straightforward setup process. It offers users multiple installation options, including direct build with Stack, Docker, or Homebrew. This flexibility facilitates accessibility across different platforms and development environments.

However, users are required to define the properties they wish to test explicitly. This feature allows for customized and focused testing scenarios, ensuring the analysis is directly relevant to the contract's intended behaviors and security assumptions. Despite the advantage of tailored testing, this approach adds a layer of complexity. Users must thoroughly understand their contract's logic and potential vulnerabilities to define meaningful test properties effectively. This necessity for manual property definition demands a higher level of engagement and expertise from the user than tools that automatically infer test cases or vulnerabilities.

\subsection{Summary}
Table \ref{tab:tools-sum} presents a comparative analysis of Oyente, Slither, Mythril, Manticore, and Echidna. The comparison is based on the number of detected vulnerabilities, the methodology used, an academic or company tool, the code level, the required solidity version, and the availability of documentation.

\begin{table}[h!]

\begin{tabular}{| p{5.5em} | p{3.1em}| p{4em} | p{3em}| p{4em} | p{3em} | }

  \hline

 & Oyente & Slither & Mythril & Manticore & Echidna \\ 
  \hline
  
Number of detected vul & 4 & 70 & 14 & 10 & - \\ 
   \hline
Methodology & Symbolic & Symbolic & Static & Symbolic & Fuzzing \\ 
   
   \hline
Code Level & EVM  & Solidity & Solidity & EVM  & Solidity \\ 
   \hline
Restrictions (Solidity) & $\leq$ 0.4 & $\geq$ 0.4 & $\geq$ 0.4 & $\geq$ 0.4 & not specified \\ 
  \hline
Documentation & Low & Medium & High & High & Medium \\ 

  \hline
\end{tabular}
\caption{Qualitative comparison of selected tools }
\label{tab:tools-sum}
\end{table}

In conclusion, there are various tools that can be used to test smart contracts and identify a wide range of vulnerabilities. Static analysis tools are still the most popular because they are simple to use compared to (a) fuzzing, which necessitates a lot of resources to set up the test environment, and (b) formal verification, which necessitates expertise in contract specification writing.

\section{Guidelines for secure smart contracts}
\label{dis}
\begin{figure}[h]
    \centering
    \includegraphics[width=0.48\textwidth]{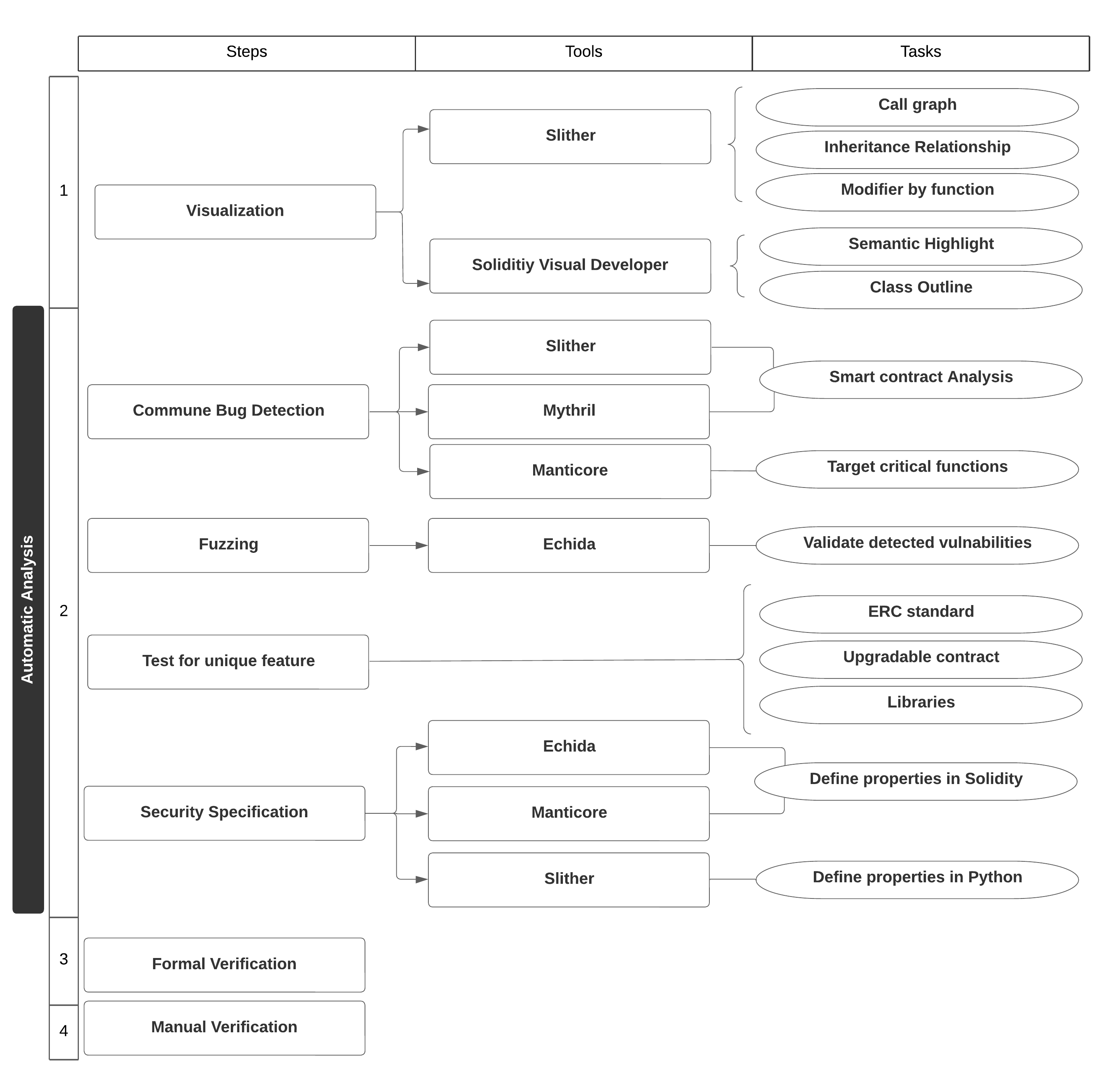}
    \caption{Smart contract auditing}
    \label{fig:auditing}
\end{figure}

Writing a secure and bug-free smart contract is crucial, as a small bug can lead to the loss of millions of dollars. In this section, we provide guidelines on how to write a secure smart contract \cite{CHECKLIST}. These steps are summarized in Figure \ref{fig:auditing}

The first step in testing a smart contract is to visualize its control flow correctly to have a global view and better understand the interactions between its different components. \emph{Slither} is an exciting tool to use during this step since it has a variety of printers that can be used to visualize call graphs, contract inheritance relationships, modifiers called by each function, etc. \emph{Solidity Visual Developer} is another interesting tool that could be integrated into Visual Studio Code \cite{Visual}. This extension helps with semantic highlighting. It also generates a detailed class outline. This step is essential to spot critical functions that must be tested in the next stage.

Afterward, the smart contract should be subjected to automatic analysis. It is best to start with common bug detection using  \emph{Mythril} and \emph{Slither}; we recommend combining tools for better results. To target a certain critical function, it is best to use the \emph{Manticore} tool as it allows the implementation of specific use cases. It is also crucial to employ dynamic analysis tools to decrease the number of false positives generated by static tools and to cover additional states. \emph{Echidna} is an interesting ethereum fuzzing tool to use \cite{echidna}. 
It is recommended to look for unique features in smart contracts. For instance, if the smart contract is an ERC token, it will be essential to use \emph{Slither's ERC Conformance} functionality to make sure that the smart contract conforms to the ERC standards put in place.

Because automated tools are unaware of the context for which the contract was produced, running more specific testing will increase the detection accuracy. To achieve this, \emph{Echidna} and \emph{Manticore} \cite{manticore-verifier} allow for the definition of security properties in solidity. Indeed, these tools help check arithmetic operations, external interactions, and standardization. Whereas \emph{Slither} allows the definition of properties with Slither Python APIs, it will enable checking for inheritance, variable dependencies, and access restrictions.

The third step will be formal verification, which ensures that the contract requirements are legitimate. This step, however, is challenging as it requires unique expertise (e.g., writing the formal specification using a specification language). Finally, the output of the preceding steps must be manually verified, and corrective actions based on patterns and mitigation techniques must be applied. After fixing the bugs, one has to redo the previous steps to increase the probability of a bug-free contract.

\section{ Related Work}
\label{related}

Several surveys covering smart contract vulnerabilities have been published in the last few years. The majority of these surveys cover one or two of the following areas: smart contract security vulnerabilities, analysis of existing tools and methodologies to detect smart contract vulnerabilities, formal specification and verification, and common design patterns and practices. 
Kushwaha et al.\cite{9762279} conducted a systematic review of Ethereum smart contract analysis tools, classifying them into static and dynamic analysis categories. They scrutinize 86 tools, discussing their methodologies, such as taint analysis, symbolic execution, and fuzzing. However, it lacks consideration for the developer community's preferences and practical scenarios where each analysis tool would be most beneficial. 
Kushwaha et al. \cite{9667515} presented a comprehensive systematic review of security vulnerabilities in Ethereum blockchain smart contracts. The paper categorizes vulnerabilities into three main root causes and seventeen sub-causes, providing in-depth insights into twenty-four specific vulnerabilities and their prevention methods, detection, and analysis tools. Despite providing a comparative analysis of various detection tools, the paper falls short in offering in-depth feedback on the effectiveness and limitations of these tools. 
Di Angelo et al. \cite{8782988} provided a state-of-the-art review of analysis tools of Ethereum smart contracts. The study is based on the actual execution of the tools. It classifies the tools based on availability, maturity level, purpose, and analysis method. However, the survey \cite{8782988} becomes obsolete (e.g., no longer maintained).
Harz et al. \cite{Harz} presented languages, paradigms, and a verification approach for smart contracts. They did not, however, go into depth about the verification tools or known vulnerabilities.
Li et al. \cite{LI2020841} discussed blockchain safety problems and proposed enhancements; however, the survey \cite{LI2020841}  is general because, for example,  it included different types of blockchain, such as Ethereum and Bitcoin.
Saad et al. \cite{Saad} investigated several attacks on various blockchain platforms and protection mechanisms. They briefly covered mitigation schemes. However, they did not cover tools to detect vulnerabilities.
Atzei et al. \cite{Atzei} presented significant vulnerabilities and attacks related to Ethereum smart contracts. However, they did not cover mitigation schemes. 
Chen et al. \cite{Chen} presented 40 types of Ethereum vulnerabilities, their causes, and 29 attacks; however, they did not cover tools to detect vulnerabilities.
Zhu et al. \cite{Zhu_2020} presented 11 smart contract vulnerabilities in-depth, along with various defenses against well-known attacks. They covered schemes to detect those vulnerabilities.
Durieux et al. \cite{Durieux} conducted an empirical review of nine automated analysis tools on 47,518 contracts; they developed a framework to analyze these tools. The analysis uses two data sets of smart contracts with tagged vulnerabilities. However, the study did not cover vulnerabilities of smart contracts and mitigation schemes. 

We conclude that existing surveys focus only on one or two areas of smart contract vulnerabilities. Furthermore, none of the studies discuss the tools based on their output format. 

\section{Conclusion}
\label{conc}
In this study, we presented detailed common vulnerabilities in Ethereum smart contracts and mitigation solutions based on patterns and standards. We have covered the most popular detection methodologies: static analysis, dynamic analysis, and formal verification. We also discussed the benefits and drawbacks of each method. In addition, we provided community-recommended vulnerability detection tools, an execution sample, and detailed feedback for each. Based on our investigation of existing methodologies and tools, we proposed a guideline with the tool(s) for each step in auditing smart contracts.

We observed that most tools detect vulnerabilities but do not provide refactoring recommendations. In future work, we aim to implement a refactoring module that proposes improvements based on the detected bugs and error traces. We also aim to study attacks related to NFTs, such as wash trading, a type of market manipulation in which attackers repeatedly buy and sell the same NFT to increase the price and trading volume.

\bibliographystyle{ieeetr}
\bibliography{jrnl}

\end{document}